\documentstyle[prl,preprint,aps]{revtex}  
\begin{document} 
\title {Structures for Data Processing in the Quantum Regime}  
\author{Simon C. Benjamin\footnote{s.benjamin@physics.ox.ac.uk}
and Neil F. Johnson} 
%
%
\address {Physics Department, Clarendon Laboratory,  Parks
Road, Oxford OX1 3PU,  England} 
%
\date{\today} 
\maketitle

\begin{abstract}  We present a novel scheme for data processing
which is well-suited for implementation at the nanometer scale.
The logic circuits comprise two-state cellular units which are
driven by externally applied updates, in contrast to earlier
proposals which relied on ground-state relaxation. The present
structures can simultaneously process many inputs and are
suitable for conventional, dissipative computing in addition to
classical reversible computing and quantum computing.


\end{abstract}

\newpage

There has been much recent interest in the topics of
computation and information processing at the nanometer scale
where quantum mechanical effects can play an important role
\cite{Feynman,NewScientist}. Theoretical design
schemes have been reported for the regimes of conventional,
classical
computation\cite{domino,lines,Obermayer,Korner,prbCA,ACApaper}
and, more recently,  quantum computation utilizing wavefunction
coherence across the entire structure\cite{deutsch,Steane,QCqdot}. The
proposed structures usually consist of networks containing a
large number of identical cellular units or `cells'. Various
cells have been suggested, from  a group of five elementary
quantum dots arranged as a square of four with one in the
center\cite{domino} through to individual quantum
dots\cite{lines,Obermayer,Korner,prbCA,QCqdot}. A drawback of such
schemes is that a rather complex network of cells may be needed
in order to reproduce the logical operation of a given,
conventional logic circuit\cite{domino}.

This letter proposes a new architecture for such cell-based
computation which can be easily generalized to reproduce any
conventional logic circuit. The architecture has many possible
physical realizations, although we will here focus on a
specific implementation where the cells consist of coupled
semiconductor quantum dots (see Ref. \cite{Obermayer} and
below). Each cell has two distinct internal states, referred to
as `0' and `1',  in the energy range of interest.  One
important requirement of the network design is that it should
be possible to change the state of a cell conditional on the
cell's current state and the states of its neighboring cells.
The cells in the network are only required to be sensitive to
the states of their nearest neighbors: we will take the term
`neighbors' to mean `nearest neighbors'. The cells are {\em
not} required to be able to distinguish one neighbor from
another. Such a requirement would reduce the range of possible
physical realizations. Moreover in the particular case of the
double quantum-dot implementation, the requirement that cells
distinguish their neighbors effectively prohibits
two-dimensional networks. It is a feature of all our designs that a cell
and its neighbors are never addressed by the same update.

The general scheme employs a number of different `types' of
cell, where `type' denotes a subset of cells with the same
energy separation between `0' and `1'. With a greater number of
types one can produce more compact designs. Conversely, using
more complex designs one can perform certain functions with only
a single cell type; this is described more fully in
Ref.\cite{longPaper}. Here we present patterns
that represent a good `trade-off' between number of cell types
and the pattern's complexity. We will employ two cell types to
produce patterns which perform certain elementary functions,
namely {\em transportation of data}, {\em fanning-out (i.e.
copying) of data}, and the {\em logical operations} XOR and
NOR. This set of components suffices to produce the cellular
equivalent of any conventional logic circuit. We then proceed
to show that a modified version of this cellular scheme is
suitable for reversible classical computing \cite{bennett,landauer} and
quantum computing. The authors have made available on the internet a
Java Applet
\cite{applet} which allows the user to follow the time
evolution of the patterns shown here, and to design
new patterns.

Figure 1 shows how cells of just two types, $\alpha$ and
$\beta$, can be arranged to produce the above-mentioned
elementary functions. The cells may have one, two or three
neighbors. Those cells having just one neighbor are referred
to as `terminators' since they maintain the same state (`0' in
all cases shown here - see Fig. 1).  In order that the patterns
carry out their data processing functions, they must be subject
to a certain repeating sequence of conditional updates. For the
conditional updates we employ the notation $\buildrel
t\rightarrow u \over  {w_v}$ to denote the following: cells of
type $w$ which are presently in state $t$ will change to state
$u$ if and only if the `field' is of strength $v$; the `field'
is defined as the number of nearest neighbors in state `1'
minus the number in state `0'. For example,
$\buildrel 1\rightarrow 0 \over {\beta_{-2}}$ indicates that
those cells of type $\beta$ whose current state is `1' are to
change their state to `0' if, and only if, two more neighbors
are in state `0' than are in state `1'. 

Figure 1(a) provides a pattern which acts as a wire. Suppose
that at some instant the arrangement of data bits is as shown
in Fig. 1(a)(i), i.e. one bit of information is represented by
the state of each  $\alpha$ cell which has two neighbors. We
use symbols such as $x_i$ to label these data bits along the
`wire'. After application of the update $\buildrel 0\rightarrow
1 \over {\beta_0}$ followed by
$\buildrel 1\rightarrow 0 \over {\alpha_0}$ the bits will all
have moved one cell to the right, so that the
$\beta$ cells with two neighbors now represent the data bits
and the
$\alpha$ cells are all in state `0'. If the updates
$\buildrel 0\rightarrow 1 \over {\alpha_{-1}}$,$\buildrel
1\rightarrow 0
\over {\beta_0}$ are now applied, the data bits will move one
cell further, onto the $\alpha$ cells with three neighbors.
Similarly
$\buildrel 0\rightarrow 1 \over {\alpha_0}$, $\buildrel
1\rightarrow 0 \over {\alpha_{-1}}$ will move the data onto the
set of
$\alpha$ cells with two neighbors, so completing the cycle
(Fig. 1(a)(ii)). By repeatedly applying the sequence
  
\noindent 
$\buildrel 0\rightarrow 1 \over {\beta_0}$,
$\buildrel 1\rightarrow 0 \over {\alpha_0}$,
$\buildrel 0\rightarrow 1 \over {\alpha_{-1}}$,
$\buildrel 1\rightarrow 0 \over {\beta_0}$,
$\buildrel 0\rightarrow 1 \over {\alpha_0}$, 
$\buildrel 1\rightarrow 0 \over {\alpha_{-1}}$

\noindent the stream of bits is caused to flow along the
`wire', progressing three cells with each repetition.

Figure 1(b) shows a pattern which performs the {\em copy} or
{\em fanout} operation: one stream of data is divided into two
identical copies. The patterns shown in Fig. 1(c),(d) and (e)
perform the logical operations XOR, NOR and NOT,
respectively. The following master sequence of updates suffices
for any and all of the patterns shown in Fig. 1:

\noindent 
$\buildrel 0\rightarrow 1 \over {\beta_0}$,
$\buildrel 1\rightarrow 0 \over {\alpha_0}$, (
$\buildrel 0\rightarrow 1 \over {\alpha_{-1}}$,
$\buildrel 0\rightarrow 1 \over {\beta_{-3}}$),(
$\buildrel 1\rightarrow 0 \over {\beta_0}$,
$\buildrel 1\rightarrow 0 \over {\beta_{-2}}$),
$\buildrel 0\rightarrow 1 \over {\alpha_0}$, (
$\buildrel 1\rightarrow 0 \over {\alpha_{-1}}$, 
$\buildrel 1\rightarrow 0 \over {\alpha_1}$,
$\buildrel 1\rightarrow 0 \over {\beta_{-1}}$).

\noindent The brackets in the above sequence enclose
sub-sequences which can be performed in any order, or
simultaneously. Repeating the above sequence will cause data to
flow through {\em any} circuit formed from the components shown
in Fig. 1. It is therefore straightforward to produce a complex
cellular pattern that performs any function for which a
conventional logic diagram is known. In Fig. 2 we provide an
example of such a circuit which performs the task of adding
binary numbers.  Figure 2(a) shows how the minor circuit known
as a `half-adder' can be built from NOT, XOR and NOR gates.
Figure 2(a) also shows the corresponding cellular half-adder
formed from the cellular components of Fig. 1. Note that in
Fig. 2 the passive `terminator' cells are drawn with dashed
borders.

The upper part of Figure 2(b) shows a novel addition circuit
formed from half-adders and XOR gates \cite{novelAdder}. In
this diagram the symbols
$A_i$ ($i=0..2$) denote the bits of a number $A$ in binary
form, i.e. $A=\sum_i A_i 2^i$, and similarly for $B$ and $T$.
When any two three-bit numbers $A$ and
$B$ are input to this circuit, the sum $T=A+B$ is output. The
`depth' of the circuit is eight gates. The lower part of Fig.
2(b) shows how the novel addition circuit can be directly
translated into cellular form by replacing each gate with the
appropriate cellular pattern from Fig.1. Note that in this
complex pattern, terminators sometimes occur with two
neighbors - despite this they still remain in state `0'
throughout.

An important feature of our cellular circuitry is the ability
of a single circuit to simultaneously process a long sequence
of inputs. Consider the XOR gate shown in Fig. 1(c). This
pattern is shown containing three independent sets of bits, 
[$x_i$,$y_i$], [$x_{i+1}$,$y_{i+1}$], XOR($x_{i+2}$,$y_{i+2}$);
the other patterns shown in Fig. 1 also contain three
independent sets of bits. In general a circuit formed from
these components with a total `depth' of $n$ cells can
simultaneously process $n$/3 sets of bits. The cellular
addition circuit shown in the lower part of Fig. 2(b) has a
depth of 45 cells, and at a given instant it will be processing
the addition of 15 separate pairs of numbers. This property may
be very advantageous for certain problems such as
multi-dimensional numerical integration in which the same
function must be applied to a vast number of inputs; provided
that the number of inputs is much greater than $n$/3, then the
time required to process each input is just one repetition of
the update sequence, regardless of the circuit depth $n$. This
property, which we refer to as `simultaneous sequential
processing', is distinct from (but could  possibly act in
addition to) the phenomenon of `quantum parallelism' mentioned
below.

So far we have discussed only conventional, irreversible
computation. We have employed two-input, one-output gates and
moreover we have used irreversible updates. These updates were
of the form $\buildrel t\rightarrow u \over  {w_v}$ where $t$
denoted the initial state and $u$ the final state; this ability
to address cells of a given state allowed us to irreversibly
dissipate information. To see this, consider a simple line of
cells ...$\alpha$$\beta$$\alpha$$\beta$$\alpha$... in which one
$\alpha$ cell is in state `1' and all the other cells are in state `0'.
Applying the update
$\buildrel 1\rightarrow 0 \over {\alpha_{-2}}$ will change the `1' to a
`0' {\em without altering any of the other cells}, i.e. we will
have dissipated information. We now introduce a reversible
update using the notation $w_v$ which means that cells of type
$w$ must NOT themselves (i.e. invert their state) if and only
if the `field' (defined above) is of strength $v$. It is clear
that we may always reverse the effect of such an update merely
by repeating it (recall that a cell and its neighbors are never addressed
by the same update). More generally we may reverse a sequence of such
updates merely by repeating them in reverse order. Figure 3(a) shows a
line of cells that can act as a wire when subject to reversible updates.
With each single bit of data represented by the states of a pair of
adjacent cells (i.e. 00 or 11), the short sequence $\beta_0$,$\alpha_0$
is sufficient to move all the bits forward along the wire by two cells.
It is interesting to note that this change could {\em not} have been
accomplished by the dissipitive updates used earlier.

It remains to show that the reversible updates suffice to
produce a general reversible computer. It is known
\cite{Feynman} that a reversible computer can be constructed
using just one type of three-input, three-output gate, for
example the Toffoli or control-control-not gate (CCNOT).
Therefore, we need only show that a CCNOT gate can be
implemented using our reversible updates. Figure
3(b) shows a CCNOT gate rendered in cellular form. Note that we
have introduced a third cell type $\gamma$. The following update
sequence \cite{betterthan21} will propagate information through
the gate:

\noindent
$\beta_0$,$\alpha_{-1}$,$\alpha_0$,$\alpha_1$,$\gamma_{-2}$,$\alpha_0$,(
$\alpha_3$,
$\alpha_1$),$\alpha_0$,$\beta_0$,$\gamma_0$,$\alpha_0$,$\beta_0$,($\alpha_1$
,
$\alpha_3$),$\beta_0$,$\gamma_{-2}$,$\alpha_1$,$\beta_0$,$\alpha_{-1}$,
$\alpha_0$.

\noindent If the simple wire structure shown in Fig. 3(a) were
subject to this sequence, the net effect would be to propagate
the data bits forward by six cells. Thus to operate a
reversible computer formed from CCNOT gates and connecting
wires, we need only repeatedly apply the above sequence of 21
updates.

The reversible classical computer can be seen as mid-point
between conventional computing and `quantum computing'
\cite{deutsch,Steane}, a relatively new paradigm that is receiving
considerable attention presently. A quantum computer (QC) would
exploit `quantum parallelism'; the input would be a
superposition of a vast number of classical inputs. Together
with the phenomena of entanglement and interference, this would
allow the QC to solve certain problems \cite{quantalgorithms}
such as factorization exponentially faster than any classical
computer. It is therefore interesting to see if the reversible
cellular architecture can be generalized to a full quantum
architecture. As noted above, a single three-bit gate such as
the CCNOT is sufficient to implement a general reversible
computer. For a general quantum computer, we would require two
additional one-bit gates \cite{shorErrorCorrect} which rotate a
cell's state by
$\pi$/2 about the {\em x}-axis and the {\em z}-axis of the
Bloch-sphere. The implementation of the CCNOT gate shown in
Fig. 3(b) remains valid when we generalize the bits $x_i$ to
`qubits' $x_i=A|0\rangle+B|1\rangle$, which would be represented
by a pair of cells as $A|00\rangle+B|11\rangle$. In Fig. 3(c) we
show a trivial way of implementing one of the
$\pi$/2 rotation gates by introducing a further cell-type,
$\delta$. Suppose that at some instant the two cells labeled
$x_i$ in Fig. 3(c) are in the state
$A|00\rangle+B|11\rangle$. Now consider the effect of the
following update sequence:

\noindent
$\beta_0$,$\alpha_{0}$,$\delta_{0}$,$\beta_0$,
$\delta_{-2}^{{\pi\over 2} , x}$,
$\alpha_{0}$,$\delta_{0}$,$\beta_0$,$\alpha_{0}$.

\noindent Immediately after the fourth update is applied, the
qubit $x_i$ will be stored on only the $\delta$ cell, whose
state will be
$A|0\rangle+B|1\rangle$. The update written
$\delta_{-2}^{{\pi\over 2} , x}$ represents a
$\pi$/2 rotation of the $\delta$ cell's state about the {\em
x}-axis of the Bloch-sphere. After this update the state of the
$\delta$ cell will be
${1\over 2}{\sqrt 2}[ (A-B)|0\rangle+(A+B)|1\rangle ]$. The
remaining updates will move this rotated qubit three cells to
the left so that it is represented by a
$\beta$,$\alpha$ pair in the state ${1\over 2}{\sqrt 2}[
(A-B)|00\rangle+(A+B)|11\rangle]$. The implementation of the
gate which rotates qubits about the {\em z}-axis would be
exactly analogous, and would employ a fifth cell type
$\epsilon$. Finally we observe that the sequence 

\noindent
$\beta_0$,$\alpha_{-1}$,$\alpha_0$,$\alpha_1$,$\gamma_{-2}$,$\alpha_0$,(
$\alpha_3$,
$\alpha_1$),$\alpha_0$,($\delta_0$,$\epsilon_0$),
$\beta_0$,($\gamma_0$,$\delta_{-2}^{{\pi\over 2} ,
x}$,$\epsilon_{-2}^{{\pi\over 2} ,
z}$),$\alpha_0$,($\delta_0$,$\epsilon_0$),$\beta_0$,($\alpha_1$
,
$\alpha_3$),$\beta_0$,$\gamma_{-2}$,$\alpha_1$,$\beta_0$,

\noindent $\alpha_{-1}$,$\alpha_0$.

\noindent will operate all of the components shown in Fig. 3
(including the {\em z}-axis rotator analogous to Fig 3(c));
hence this sequence would drive a general QC formed from these
components \cite{betterthan21}.

Finally we discuss a physical realization of the cell,  i.e. a
bistable double quantum-dot driven through its internal states
by laser pulses \cite{Obermayer}. The beam would encompass a
broad area and update all cells that respond to its frequency; a
single tunable laser could therefore drive an entire computer.
The potential for such structures as a realization of a
cellular automata has been considered in some detail in Ref.
\cite{Obermayer} and will only be briefly summarized here. We
will consider a cell with two neighbors since the
generalization to three neighbors is straightforward. We use
the notation x-{\bf y}-z to refer to a cell in state y whose
left and right neighbors are in states x and z respectively. 
The cell consists of  a coupled pair of quantum dots; the
low-lying single particle states are those for which the
electron is localized on one or other of the dots. The lowest
energy state within each of the two localizations are the
physical representations of `1' and `0'. One dot is assumed to
be slightly smaller than the other, so that the two
localizations are non-degenerate. The energy difference and 
wavefunction overlap are very small so that the rate of
spontaneous decay from one dot to the other is much longer than
the total computation time. The circuitry is built up from such
cells simply by producing the appropriate pattern of them.
There is no tunneling allowed between the cells; they `feel'
the states of their neighbors via the Coulomb interaction. The
Coulomb repulsion is greater for a pair of cells in the same
state than for a pair in opposite states; the energy difference
has a
$r^{-3}$ dipole-like form where $r$ is the cell-cell
separation. If the distance to a cell's neighbor on the right
is equal to the distance to its left neighbor, then the two
stable energy levels for an isolated cell split into six levels
(see Fig. 4). The idea of having more than one type of cell is
realized by using different sizes of double-dot. The size
difference shifts the double-dot's energy levels and thus makes
it possible to address one type at a time. In order to produce
the desired updates of the cells, a third transient state is
employed \cite{Obermayer}. This is a single-particle state in
which the electron probability distribution is spread over both
dots. The transient state spontaneously decays very quickly
into one of the stable states. As shown in Fig. 3, an
irreversible update is produced by pumping the system with
light of a frequency that will excite cells of one size from a
given stable state (say 1-{\bf 0}-0) into the transient state. 
The cell may decay from the transient state into either the
original state (1-{\bf 0}-0) or the flip state (1-{\bf 1}-0).
However, if the former occurs the electron will be re-excited
by the pump, so that we can  flip the state with any desired
certainty ($<1$) simply by using a pulse of sufficient
duration. Note that the frequency width of the pulse must be
sufficiently narrow to excite from only one of the levels shown
in Fig. 4, yet sufficiently broad to cover the sub-splitting
(shown shaded gray) due to non-neighboring cells. For a
reversible update, $\pi$-pulses can be used to switch
states.

The authors wish to thank Artur Ekert and Wim van Dam for useful
discussions. This work was funded by an EPSRC Grant for Photonic
Materials.

\newpage \centerline{\bf Figure Captions}

\bigskip

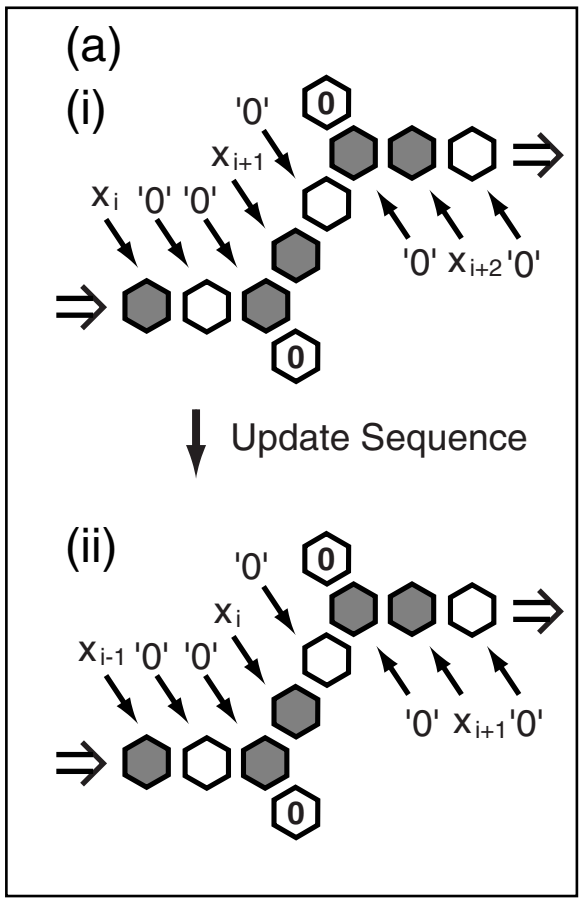
\noindent Figure 1:  A set of cellular circuit components: (a)
a wire before (i) and after (ii) the application of the update
sequence. (b) Fan-out. (c) XOR gate. (d) NOR gate. (e) NOT gate.

\bigskip 

\noindent Figure 2:  (a) The cellular form of a `half-adder'
circuit. (b) Top: a novel addition circuit. Bottom: the
addition circuit rendered in cellular form.
\bigskip

\noindent Figure 3:  (a) A `wire' suitable for reversible
updates. (b) A Toffoli or control-control-not gate. (c) A
rotation gate required for quantum computing.
\bigskip

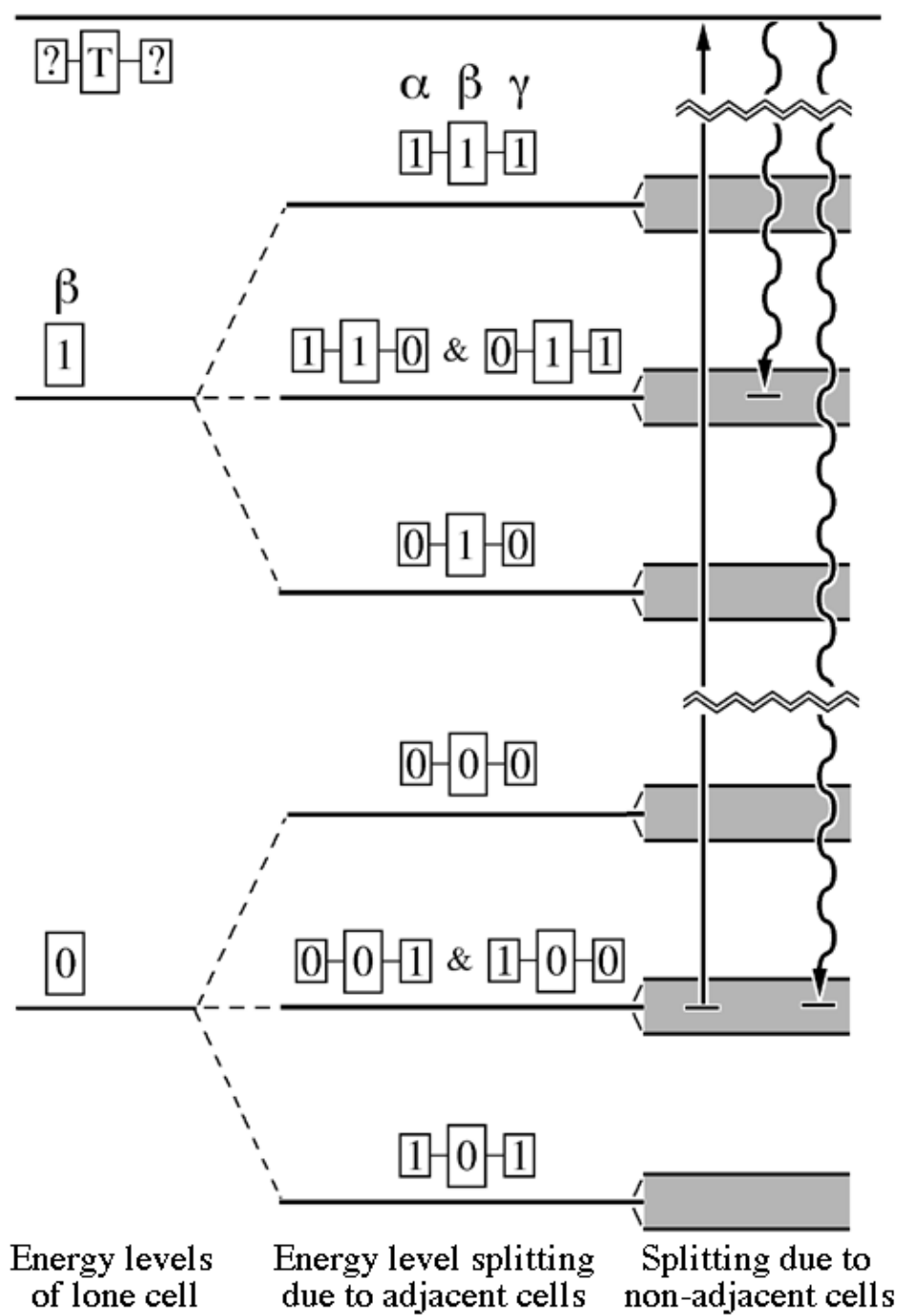
\noindent Figure 4:  The splitting of the energy levels of a
two-neighbor cell as the states of its neighbors are
specified. The shaded regions on the far right represent
splitting due to the states of non-adjacent cells.


\begin{thebibliography}{99}
\bibitem{novelAdder} We have designed this circuit to be
particularly well suited for rendering in cellular form. We
could equally well have exhibited a cellular form of the well
known `ripple carry' adder, but that adder would require twice
the depth (measured in half-adders). Note that our novel adding
circuit can easily be generalized to add two $n$ bit numbers,
the depth would then be $n$ half-adders.
\bibitem{betterthan21} There may exist shorter sequences that
perform the same function; the quoted sequence is the shortest
that the authors have found so far.
\end{thebibliography}
 
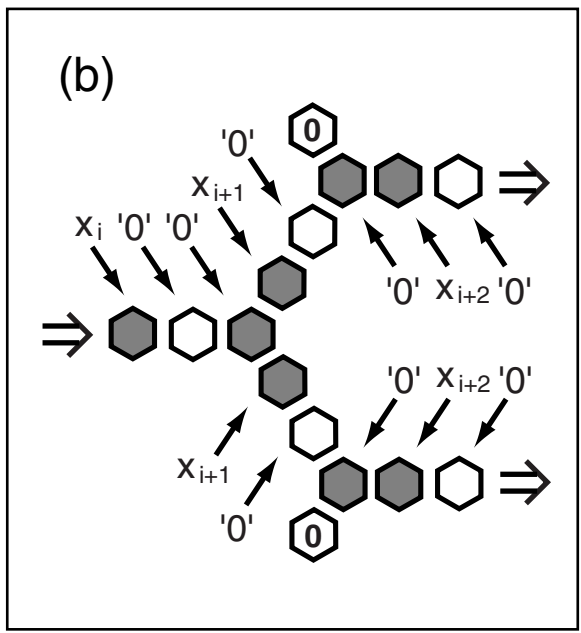
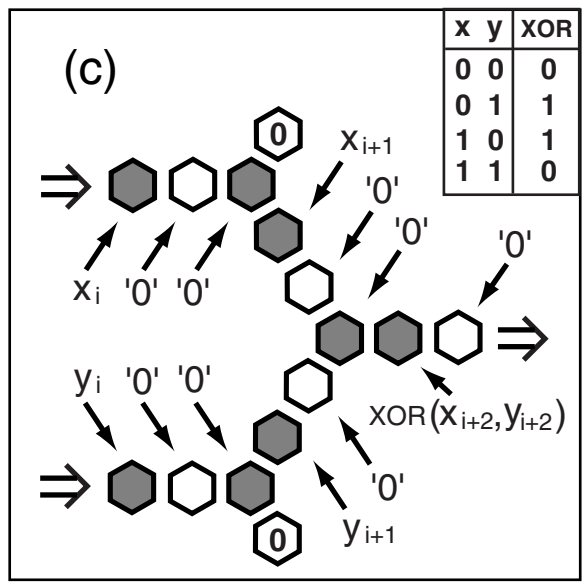
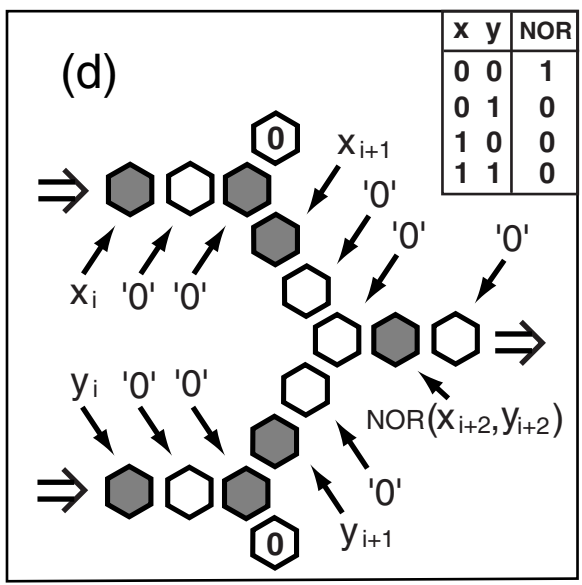
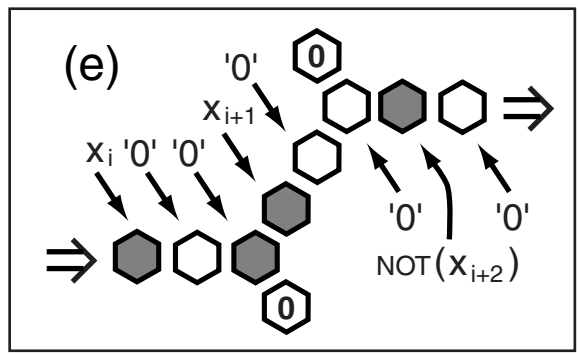
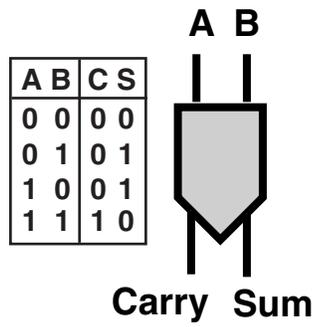
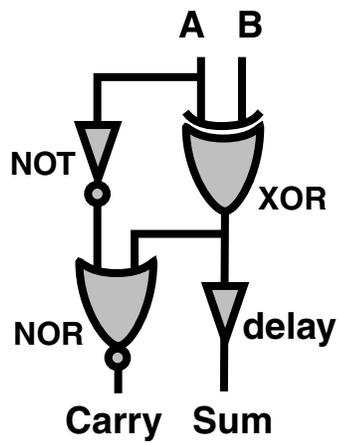
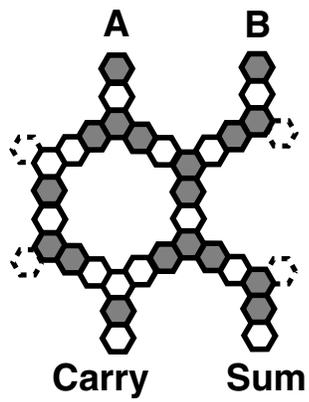
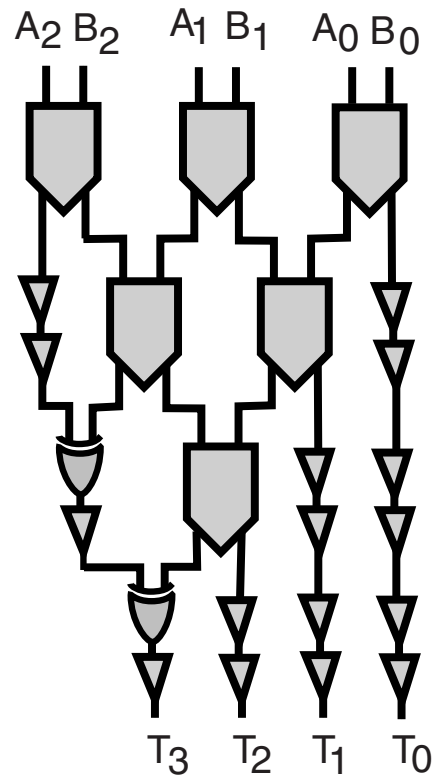
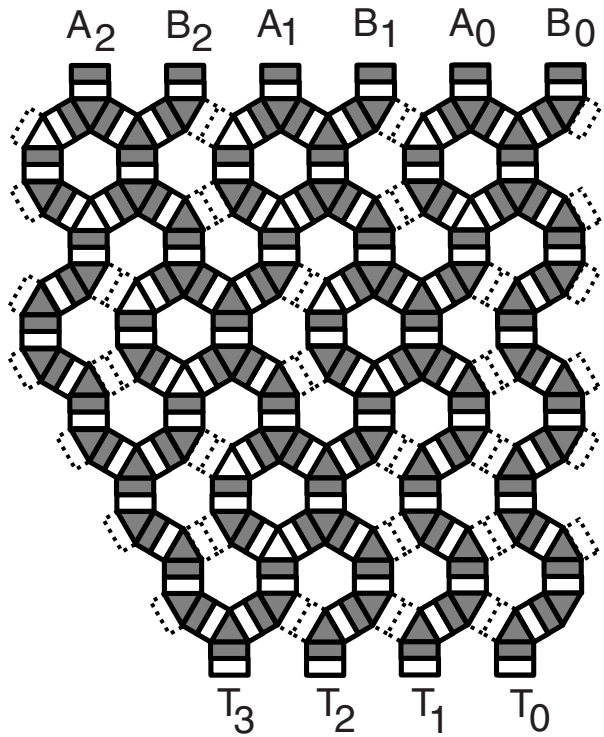
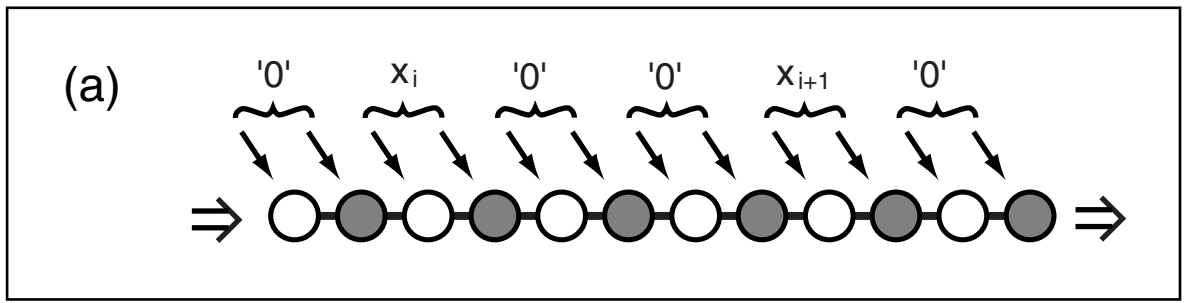
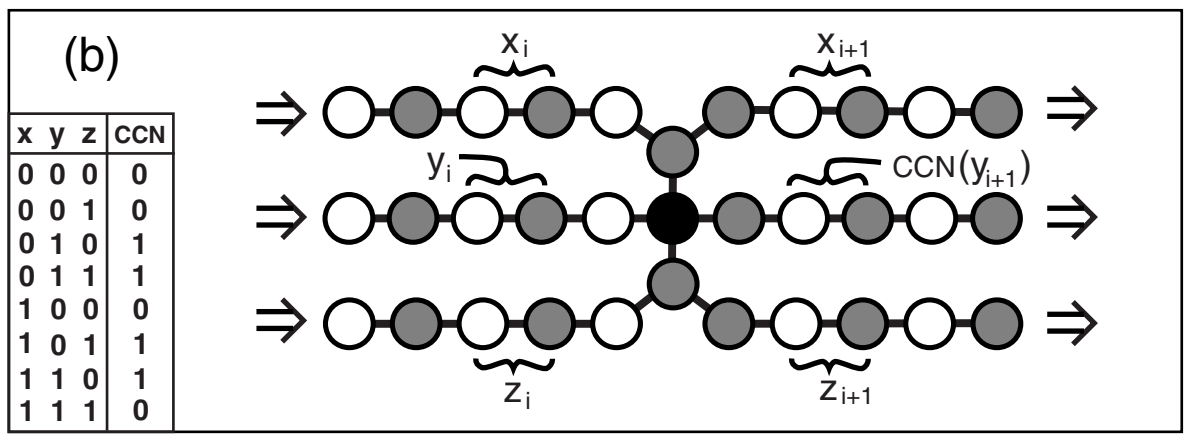
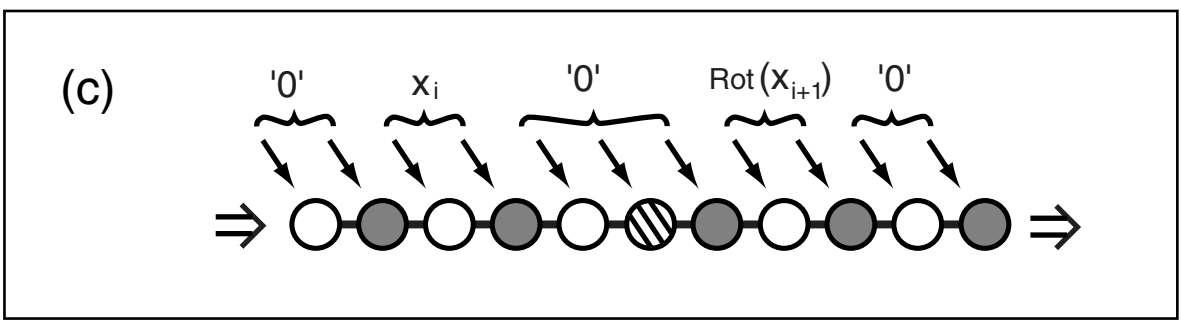
\end{document}